**ICMIEE22-045**

# Characteristics of a Nickel Vanadium redox flow battery based on Comsol

*Anupam Saha*[1,*], *Shinthia Binte Eskender*[2]
[1,2] Department of Materials Science and Engineering, Khulna University of Engineering & Technology, Khulna-9203, BANGLADESH

**ABSTRACT**
The overpotential, dissociation rate, electrode potential distributions and current density are suggested in this study to analyze the Nickel Vanadium Redox Flow Battery (NVRFB). Due to its large capacity and ecofriendly properties, NVRFB may be a viable option in the present state of energy constraint and environmental pollution. Due to their low cost and high energy density, nickel-based flow batteries have gained popularity. This study demonstrates that the Ni2+/Ni+ and V5+/V4+ ions have a higher rate of dissociation at the membrane and a lower rate at the inlet, where the electrolyte flow velocity is greater; Because the membrane undergoes more oxidation-reduction reactions, the electrolyte flow rate is critical in the redox flow cell; Additionally, we see that when electrode thickness is reduced, current density and electrode potential increase while overpotential decreases; the model's equations are solved using the finite-element method in the COMSOL Multiphysics program. An electrolyte-electrode interface connection is used to simulate the reaction. The dissociation rate indicates that the oxidation-reduction process happens at a lower membrane potential. Improving the electrolyte flow rate enhances battery performance. Compression of the electrodes enhances conductivity and battery performance.

## 1. Introduction

When it comes to bridging the gap between current technology and a clean sustainable future in grid reliability and utilization, it is anticipated that energy-storage systems will play a more prominent role. This is due to the growing need to seamlessly integrate renewable energy with the existing electricity grid, which is itself developing into a more intellectual, efficient, and suitable electrical power system, it is anticipated that energy-storage systems will play a more prominent role. Redox flow renewable technology is a cutting-edge method to offering a well-balanced answer to today's complex problems and issues [1]. It has several properties such as power output as well as capacity which may be designed in a number of ways; improved energy efficiency; comparatively higher dependability and stability; power - efficient; and a longer battery life span [2]. But they are not ideal for tiny mobile equipment such as electric vehicles since their energy density is around one-tenth that of other varieties, including lithium-ion batteries.

One possibility is to employ a Nickel Vanadium Flow Battery, which has a greater charge density. It is a device for linking many physics using a complex coupling method. When it comes to the effectiveness of the Nickel Vanadium Flow Battery, factors such as the electrochemical reaction, the battery's construction, mass transfer technique, and distribution of reaction area all play a significant role. Researches have demonstrated that nickel ions-based batteries have high conductivity and also high-water solubility, both of which are beneficial for increasing the electroactive area utilization. It is possible that all this is responsible for the spectacular results of electrospinning electrodes.

A number of different approaches have been developed and used to increase the performance of the redox flow battery [3], [4]. Li et al [6] developed an AC impedance model that failed to correctly regenerate the battery's value of the current and outer voltage during the charging and discharging processes because they did not take into account the modification with in battery's parameters during charging and discharging processes. A nickel-iron battery model was proposed by C. Chakkrabharty [5] to show the properties and limitations of the ions particle charge transfer coupling and the electrochemical reaction of Nickel-Iron Redox Flow Battery . Jie Cheng studied and modeled a nickel zinc battery and analyzed the effects of the ions. This battery showed high columbic efficiency and energy efficiency [6]. For the modeling of the quality particle charge transfer coupling and the electrochemical response of a Vanadium Redox Flow Battery, Yang et al [7] developed and suggested a 2D steady-state model (VRFB). The impacts of the initial vanadium ion concentration, the initial H+ ion concentration, and the current density are the primary focus of this study. According to the researcher, the diverse impacts of electrolyte flow rate on electrodes were investigated, and the effects on the ohmic loss generated by the electrolyte and how it may be reduced were examined. Various electrolyte flow rates have an effect on the concentration distribution, compressing electrodes has an effect on changing characteristics, and different electrode temperatures have an effect on electrode

* Corresponding author. Tel.: +88-01798281181
E-mail addresses: Anupam.kuet.mse@gmail.com

voltage based on the COMSOL software by following principles of VRFB [8].

There has been the depth analysis of the stack electrode process mechanism to analyze the electrode material impact and characteristics of electrochemical reaction on the performance of the Titanium manganese battery. In that model the principles of Titanium Manganese Redox Flow Battery (TMRFB) were followed and simulated. Then the model is used to optimize performance of the battery.

However, there is a scarcity of studies concentrating on the in-depth examination of the stack electrode process mechanism in order to determine the electrode material's influence and the electrochemical reaction characteristics on the battery's performance. Following the concepts of the Nickel Vanadium Redox Flow Battery (NVRFB), the Nernst equation, and electrodynamics, this paper develops a COMSOL-based NVRFB simulation model. The model is then utilized to maximize performance in order to enhance programs, stack settings, and test under various scenarios.

## 2. Theory of NVRFB

The fundamental operation of NVRFB is fairly similar to that of a traditional battery charging-discharging system. Nickel ions in a variety of oxidation states serve as the active material for the negative electrode and are kept in the negative electrolyte container. Vanadium ions in various oxidation states are employed as the positive electrode material and are kept in a positive electrolyte container. A membrane is utilized to isolate the positive and negative electrodes from one another. In the process of electrolyte flowing over the electrode surface, an electrochemical reaction happens, and chemical energy is converted into electric energy. After that, the current is collected and transmitted via the bipolar plate circuit [9].As a result of protons exchanging between the negative and positive electrodes, the reduction of $Ni^{2+}$ to $Ni^{+}$ happens at the negative electrode and the oxidation of $V^{4+}$ to $V^{5+}$ occurs at the positive electrode during the charging process. The schematic diagram of the NVRFB is seen in Figure 1 [10].

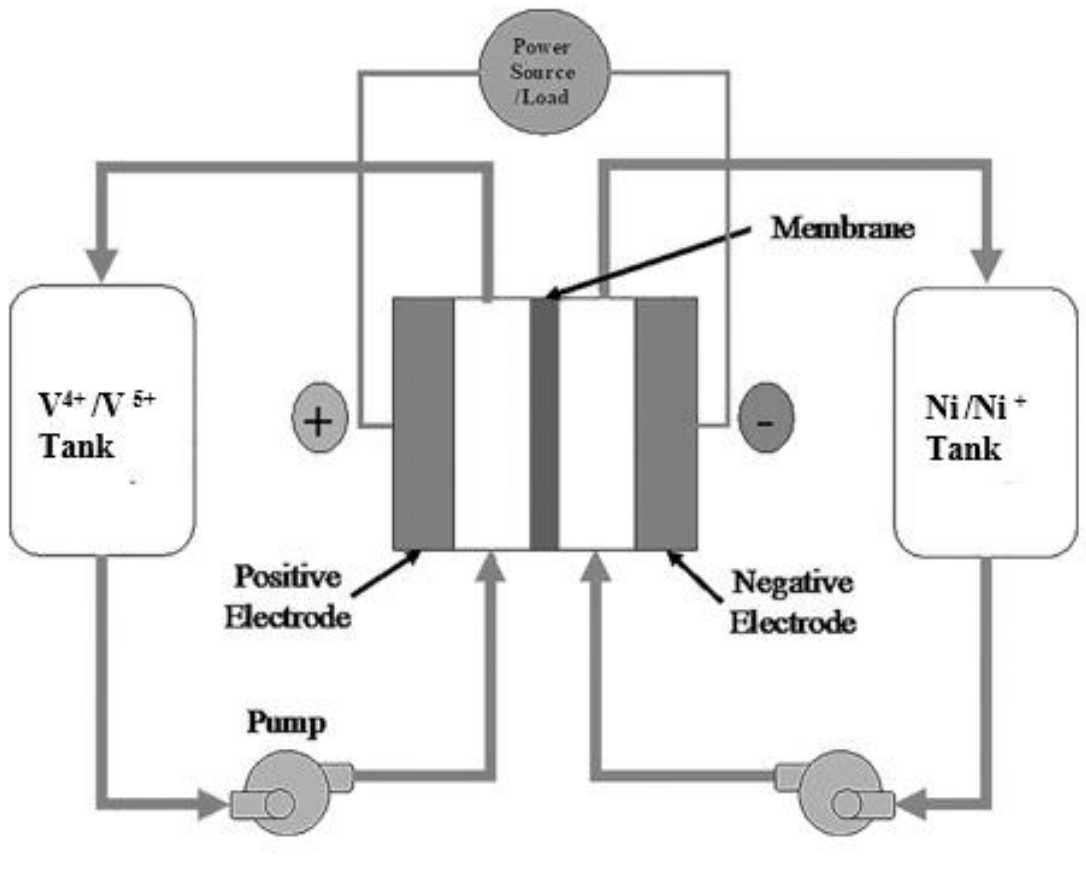

**Fig.1** The Schematic diagram of NVRFB.

the reactions between positive and negative electrodes that respond electrochemically are stated as follows:

Positive electrode reaction:

$$VO_2^{+} + 2H^{+} + e^{-} \leftrightarrow VO^{2+} + H_2O \quad (1)$$

(E= 1.004 V)

Negative electrode reaction:

$$Ni \leftrightarrow Ni^{2+} + 2e^{-} \quad (2)$$

(E= -0.24 V)

Cell reaction:

$$VO^{2+} + H_2O + Ni^{2+} \leftrightarrow VO_2^{+} + Ni + 2H^{+} \quad (3)$$

(E= 1.244 V)

## 3. Mechanism of NVRFB

It is presented in this study that a 2D steady-state model of the NVRFB, which is focused on the battery's operating mechanism, is developed. There is also a concern for concentration fields, dissociation fields, electric fields, and electrochemical reactions. The Nernst–Planck equation was used to predict the mass movement of solution species. Electrolyte flow was modeled using the Nernst-Einstein mobility relationship. NVRFB's response process is thoroughly understood, based on the influence of material qualities, the geometrical factors, and stack performance under various operating situations. A electrolyte-electrode interface coupling is utilized to simulate the reaction in a computer simulation model. Using a species interface that responds to flow concentration, we can mimic the flow and mass movement.

### 3.1 Computational Procedure

It is taken into mind that there are five components in the computational model: the positive electrode, the negative electrode, the ion exchange membrane, the positive electrolyte, and the negative electrolyte, which are all represented in Fig. In this design, the bipolar plate was taken into account. The governing equations for each component have been developed and will be presented in detail.

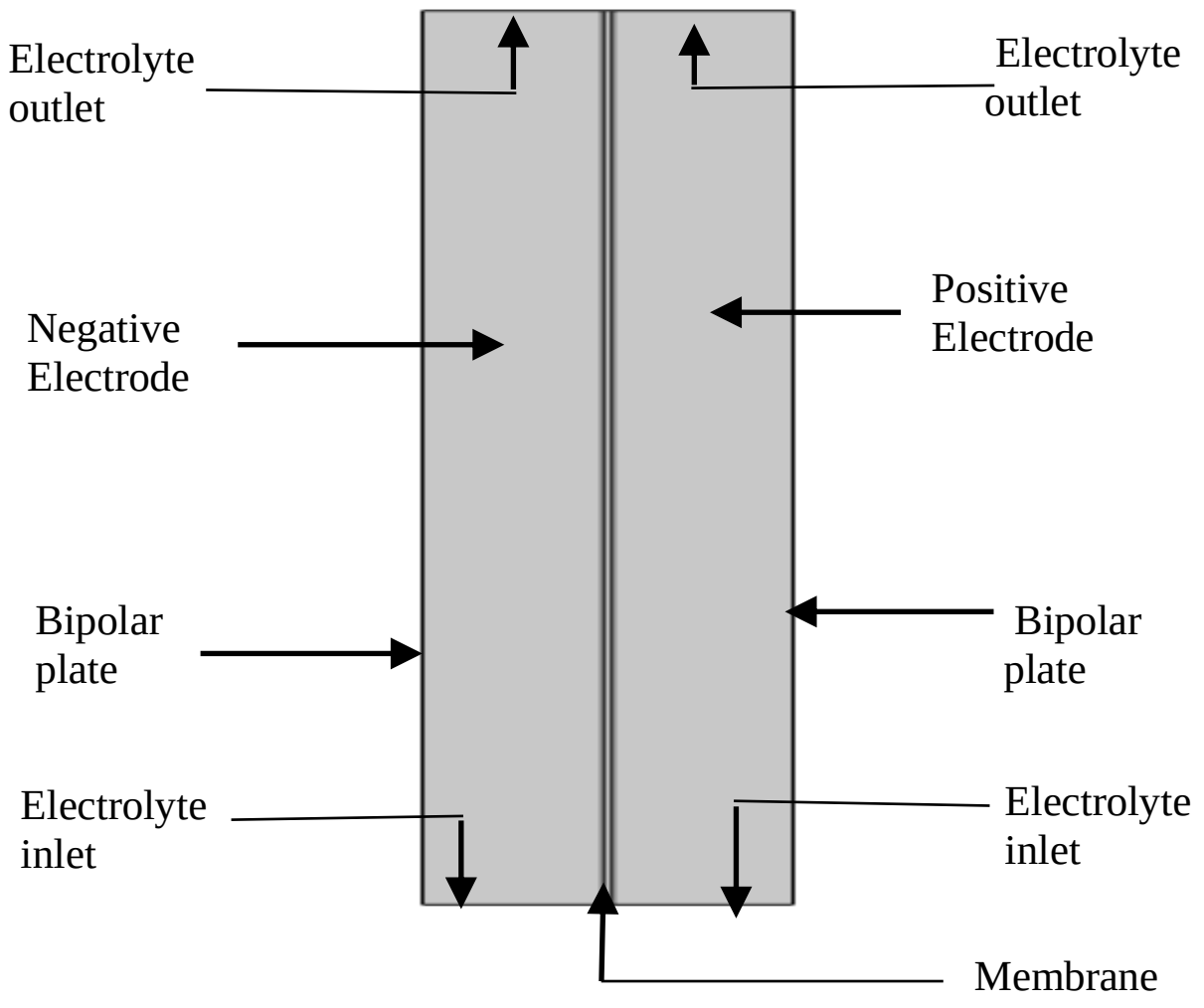

**Fig. 2** 2D Schematic Diagram

3.2 Modeling of Electrodes

Using Nernst equations, the electrode equilibrium potential can be determined:

$$E_{neg} = E_{0,neg} + \frac{RT}{F} ln\left\{\frac{\alpha_{Ni}}{\alpha_{Ni^{2+}}}\right\} \quad (4)$$

$$E_{pos} = E_{0,pos} + \frac{RT}{F} ln\left\{\frac{\alpha_{V^{4+}} \ .\alpha^2_{H^+}}{\alpha_{V^{4+}}}\right\} \quad (5)$$

The reference potential for the negative electrode reaction is $E_{0,neg}$ , whereas the reference potential for the positive electrode reaction is $E_{0,pos}$. R denotes the molar gas constant, T is the temperature, and F denotes Faraday's constant. Kinetic expressions of the Butler-Volmer type are used to explain the reactions between positive and negative electrodes:

$$i_{neg} = A.i_{0,neg}\left[e^{\frac{(1-\alpha_{neg})F\eta_{neg}}{RT}} - e^{\frac{(-\alpha_{neg})F\eta_{neg}}{RT}}\right] \quad (6)$$

$$i_{0,neg} = Fk_{neg}(\alpha_{Ni})^{1-\alpha_{neg}}.(\alpha_{Ni^{2+}})^{\alpha_{neg}} \quad (7)$$

$$i_{pos} = A.i_{0,pos}\left[e^{\frac{(1-\alpha_{pos})F\eta_{pos}}{RT}} - e^{\frac{(-\alpha_{pos})F\eta_{pos}}{RT}}\right] \quad (8)$$

$$i_{0,pos} = Fk_{pos}(\alpha_{V^{5+}})^{1-\alpha_{pos}}.(\alpha_{V^{4+}})^{\alpha_{pos}} \quad (9)$$

The negative and positive transfer coefficients are represented by αneg and αpos, respectively, while the negative reaction rate and positive reaction rate constants are denoted by $k_{neg}$ and $k_{pos}$, respectively.

Positive and negative electrode overpotentials are shown in the following:

$$\eta_{neg} = \phi_{s,neg} - \phi_{l,neg} - E_{neg} \quad (10)$$

$$\eta_{pos} = \phi_{s,pos} - \phi_{l,pos} - E_{pos} \quad (11)$$

Where the $\varphi_s$= the electrode potential, and

$\varphi_l$ = the electrolyte potential.

The porous electrode equation of current is driven from Ohm's law:

$$i_s = -\sigma_s^{eff}\nabla\phi_s \quad (12)$$

The effective conductivity $\sigma_s^{eff}$ is driven from Brueggemann correction equation. From this equation, it can be seen porous media conductivity is related to the solid materials porosity ε,

$$\sigma_s^{eff} = (1-\varepsilon)^{3/2}\sigma_s \quad (13)$$

When the thickness of the electrode is compressed, changes occur in the porosity ε of the electrode.

$$\varepsilon = 1 - \frac{d_0}{d}(1-\varepsilon_0) \quad (14)$$

Where, the $\varepsilon_0$ = porosity of the electrode,

$d_0$ = electrode thickness before compression and

$d$= electrode thickness after compression.

With constant actual surface area, electrode surface area will be modified. And it will affect the electrode's specific surface area, $\alpha$

$$\alpha = \alpha_0\frac{d_0}{d} \quad (15)$$

3.3 Dissociation of $H_2SO_4$

The following processes are followed to dissociate sulfuric acid:

$$H_2SO_4 \leftrightarrow H^+ + HSO_4^- \quad (16)$$

$$HSO_4^- \leftrightarrow H^+ + SO_4^{2-} \quad (17)$$

$r_d$ describes the degree of dissociation:

$$r_d = k_d\left(\frac{\alpha_{H^+} - \alpha_{HSO_4^-}}{\alpha_{H^+} + \alpha_{HSO_4^-}} - \beta\right) \quad (18)$$

Where $k_d$ defines $HSO_4^-$ dissociation rate constant and β refers to the dissociation level. The $H^+$ concentration in the electrolyte is affected by the dissociation of sulphuric acid, hence $r_d$ is an important parameter.

3.4 Current Density Determination

The Nernst-Planck equation helps to drive the molar flux $N_i$ of ions.

$$N_i = -D_i.\nabla_{c_i} - Z_i.u_{mob,i}.F.c_i.\nabla_{\phi_l} + c_i.u \quad (19)$$

In this equation, diffusion flux is known from the 1st term, where $D_i$ denotes diffusion coefficient and the 2nd term refers to migration term. Migration term is calculated by the electrostatic charge of the ion $Z_i$ and mobility $u_{mob,i}$. 3rd term denotes as convection term where u refers to fluid velocity.

Here current density is the multiplication of total molar flux and the number of charges of the material.

$$i_l = F.\sum_{i=1}^{n} Z_i\left(-D_i.\nabla_{c_i} - Z_i.u_{mob,i}.F.c_i.\nabla_{\phi_l}\right) \quad (20)$$

3.5 Selection of Boundary

In this model, the boundary conditions of the second current distribution interface and the boundary conditions of the tertiary current distribution interface are constructed.

1st requirement for the selection of boundary is that the electrolyte current density should be the same as the membrane current density.

$$\vec{n}.\overrightarrow{i_{l,e}} = \vec{n}.\overrightarrow{i_{l,m}} \quad (21)$$

Here, electrolyte current density = $i_{l,e}$, and membrane current density = $i_{l,m}$.

The membrane current is related to proton flux.

$$n.N_{+,e} = n.\frac{i_{l,m}}{F} \quad (22)$$

In the 2nd current distribution interface, the electrolyte potential limit is set up:

$$\phi_{l,m} = \phi_{l,e} + \frac{RT}{F} ln\left\{\frac{\alpha_{+,m}}{\alpha_{+,e}}\right\} \quad (23)$$

3.6 Computational Setting

This battery model was built using COMSOL, a finite element analysis program. This program was used to solve the model equations. Setting up the essential parameters, global definitions, model input, component and research physics was required. The standard size was followed in COMSOL to create a mesh with 4616 components. The mesh density was greater closer to the membrane, where the major reaction occurs. Parameter settings is shown in Table 1.

**Table 1** Parameters for model setup.

| Parameter | Value |
|---|---|
| Height of the cell (H) | 0.045 m |
| Depth of the cell (W) | 0.0315 m |
| The thickness of electrode (d) | 0.0042 m |
| Temperature (T) | 298 K |
| Negative diffusion Coefficient ($D_{neg}$) | $6.5\times10^{-10}$ $m^2/s$ |
| Positive diffusion Coefficient ($D_{pos}$) | $4\times10^{-10}$ $m^2/s$ |
| Specific area (*a*) | $3.5\times10^5$ |
| Positive electrode standard potential ($E_{0,\ pos}$) | 1.004 V |
| Negative electrode standard potential ($E_{0,\ neg}$) | -0.24 V |
| Positive electrode transfer coefficient ($\alpha_{pos}$) | 0.55 |
| Negative electrode transfer coefficient ($\alpha_{neg}$) | 0.38 |
| Positive electrode rate constant ($k_{pos}$) | $2.5\times10^{-8}$ m/s |
| Negative electrode rate constant ($k_{neg}$) | $4.624\times10^{-6}$ m/s |
| Initial concentration of Ni ($C_{Ni}$) | 118 mol $m^3$ |
| Initial concentration of $Ni^{2+}$ ($C_{Ni^{2+}}$) | 734 mol $m^3$ |
| Initial concentration of $V^{4+}$ ($C_{V^{4+}}$) | 734 mol $m^3$ |
| Initial concentration of $V^{5+}$ ($C_{V^{5+}}$) | 118 mol $m^3$ |

## 4. Result and Discussion

The green gradient in figure 3 represents a dissociation rate ranging from -10 to -5, whereas the orange gradient represents a dissociation rate larger than -5. The surface dissociation rate of $V^{5+}/V^{4+}$ and $Ni^{2+}/Ni^{+}$ ions is greater at the membrane and lower at the inlet where the electrolyte flow velocity is higher. It demonstrates that when velocity increases, the reaction rate of the ions decreases and more potential is required to activate the reactions. However, distant from the center of the flow, where velocity is slower, the ion reaction rate is faster, and less potential is required to activate the reactions. With the optimized flow rate of the electrolyte, a greater electrochemical reaction occurs at the membrane. The model precisely represents the internal state of the cell during operation, demonstrating that the design of the flow field is critical to the battery's optimization.

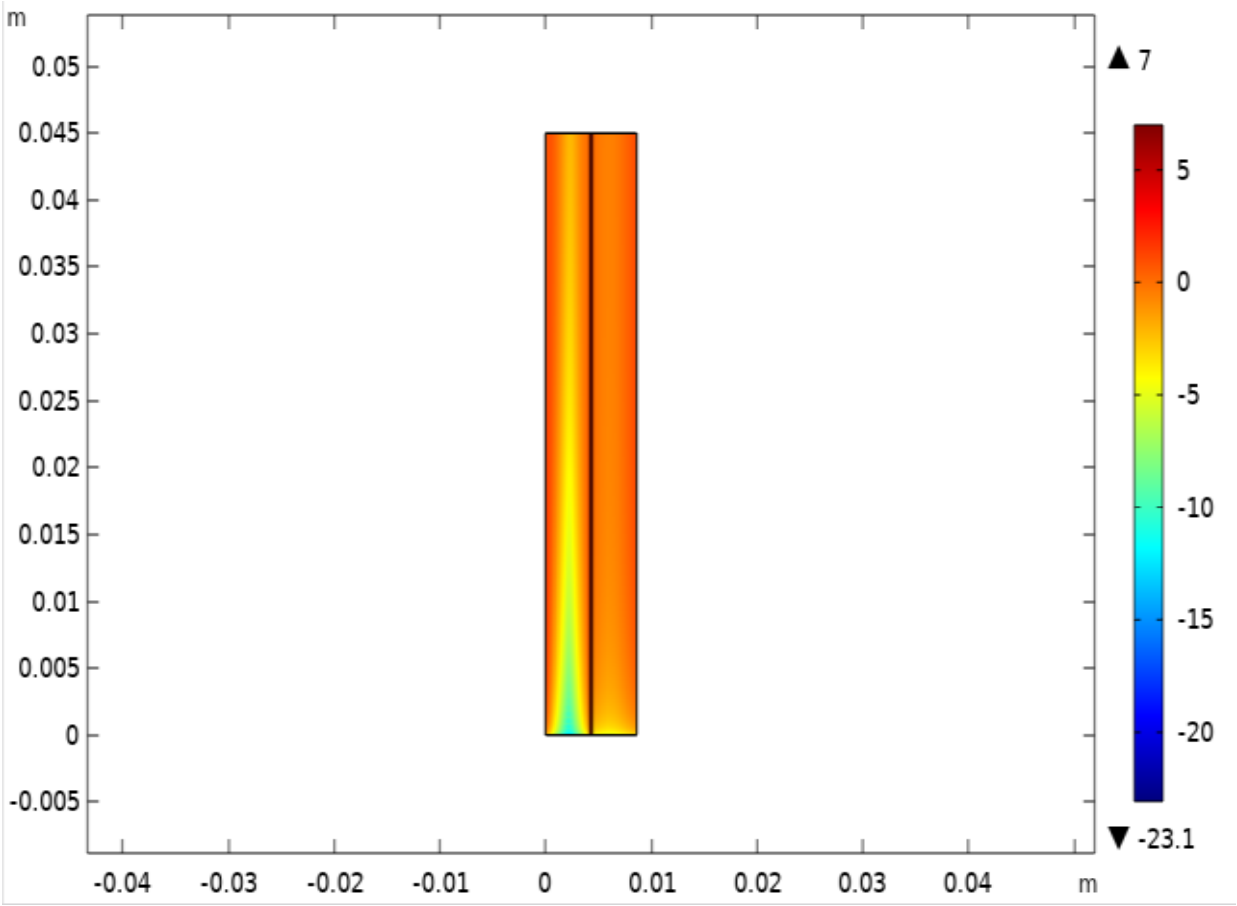

**Fig.3** Dissociation rate of $V^{5+}/V^{4+}$ *and* $Ni^{2+}/Ni^{+}$

From figure 4 to figure 7, the dissociation rate is found to be lower at the maximum flow rate and gradually increases. The flow rate of the electrolyte plays a vital role in the redox flow cell.

In figure 4 the electrolyte flow rate is 30 ml/min, the dissociation rate of $V^{5+}/V^{4+}$ ions at the middle of flow are -6.513 $s^{-1}$ and it increases fast towards the membrane. At the membrane, the rate is 0.2377 $s^{-1}$.

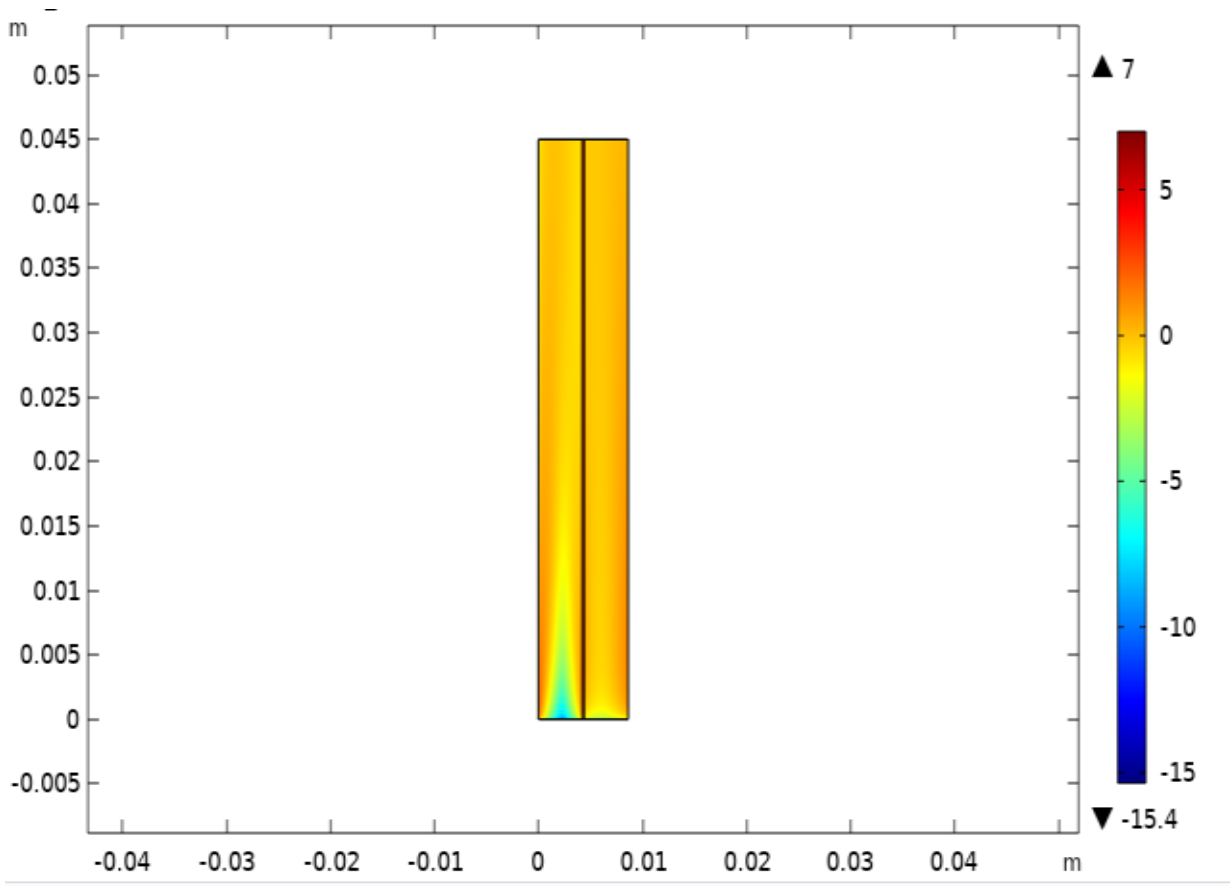

**Fig.4** Dissociation rate at flow rate 30 ml/min.

In figure 5, at a flow rate of 90 ml/min the dissociation rate increases slowly. At the middle of inlet velocity, the value is -8.376 $s^{-1}$ and slowly increases to 2.0608 $s^{-1}$ at the membrane.

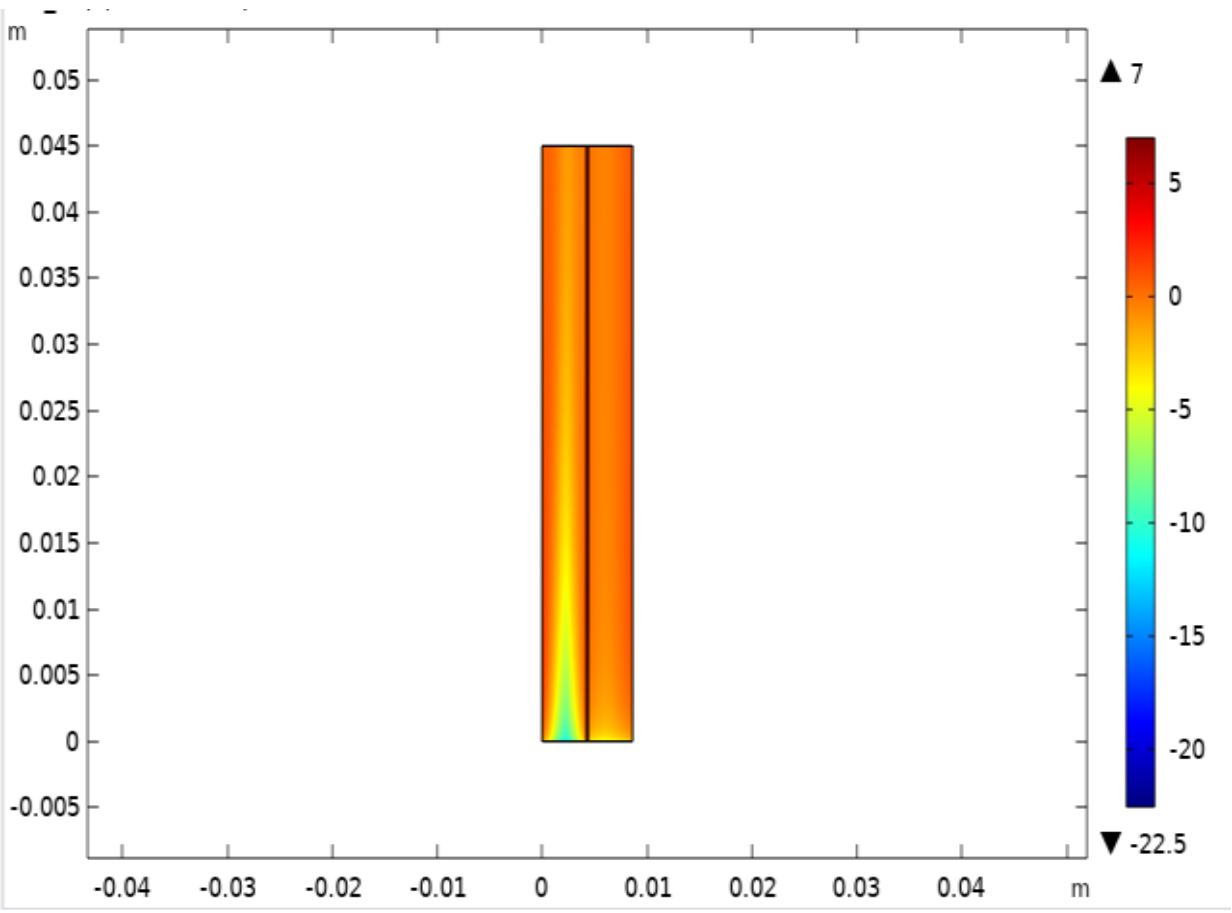

**Fig. 5** Dissociation rate at flow rate 90 ml/min.

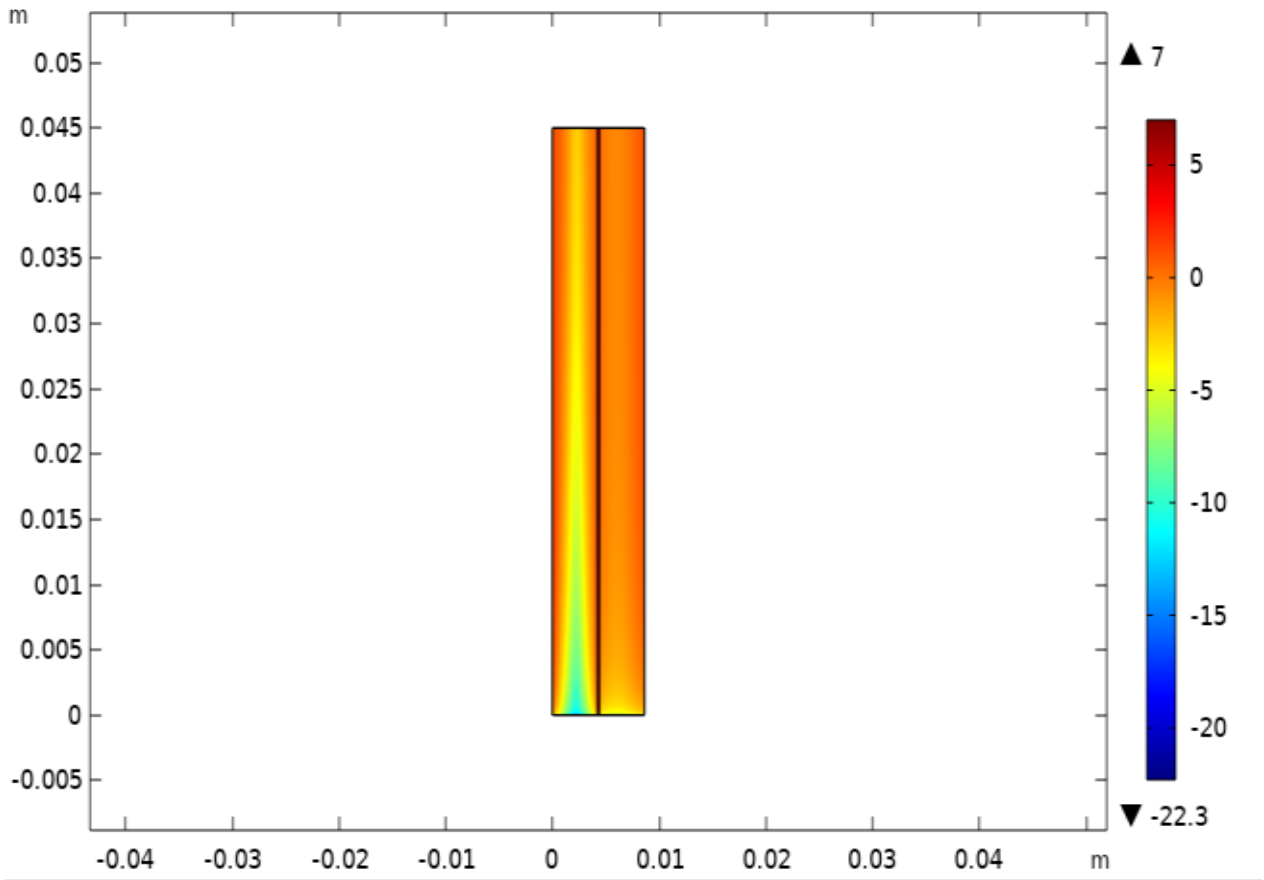

**Fig. 6** Dissociation rate at a flow rate of 180 ml/min.

The sky-blue color gradient in figure 6 represents a dissociation rate ranging from -10 to -15. When the flow rate of the electrolyte is increased, the ions do not have enough time to react and dissociate less. At an electrolyte flow rate of 180 ml/min, the dissociation rate is lower for a larger region. And it steadily climbs to 3.248 s-1 near the membrane. The rate of dissociation from the inlet to the membrane is faster in Ni-V cells.

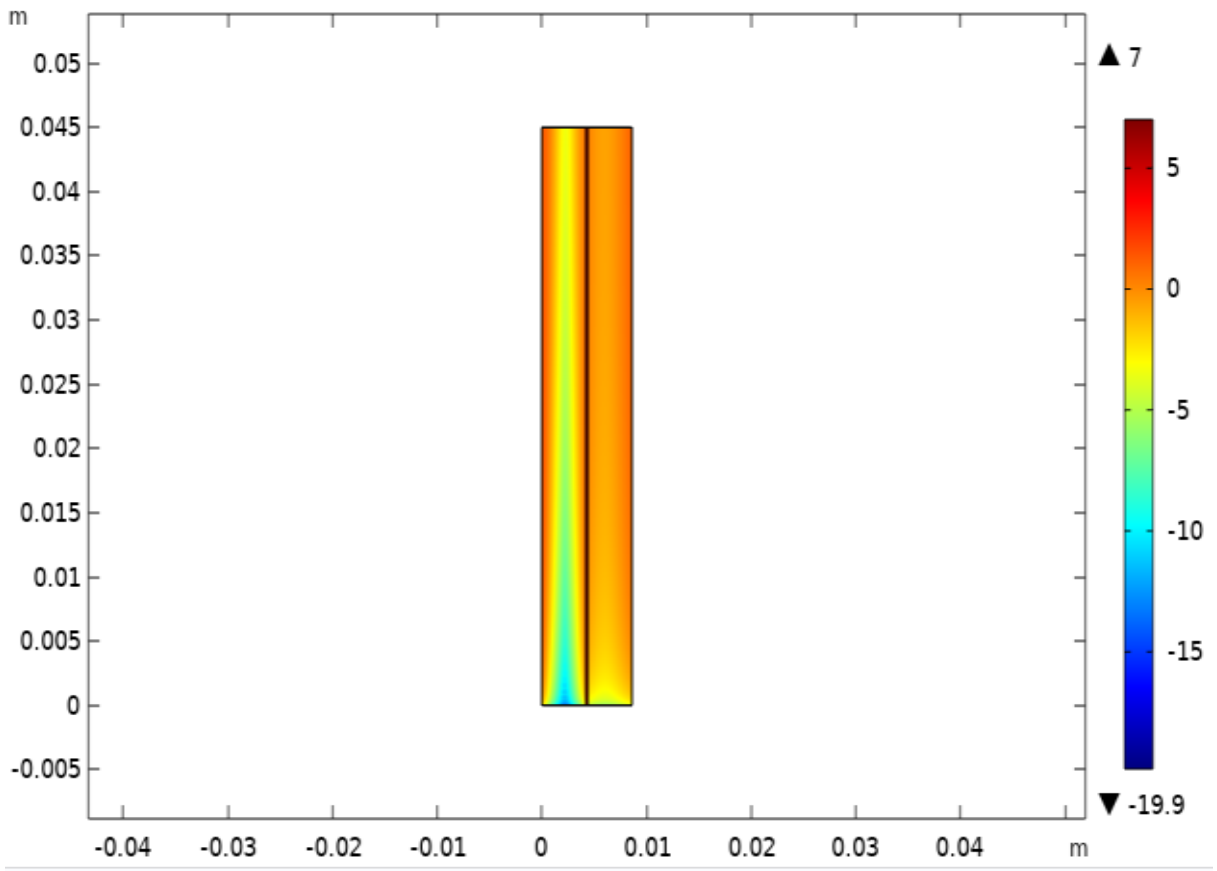

**Fig. 7** Dissociation rate at a flow rate of 250 ml/min.

Figure 7 represents the dissociation rate at electrolyte flow rate 250 ml/min. At this flow rate the dissociation rate at the inlet is ranging from -10 to -15 $s^{-1}$. The dissociation rate increases very slowly this time and reaches to -1.475 $s^{-1}$.

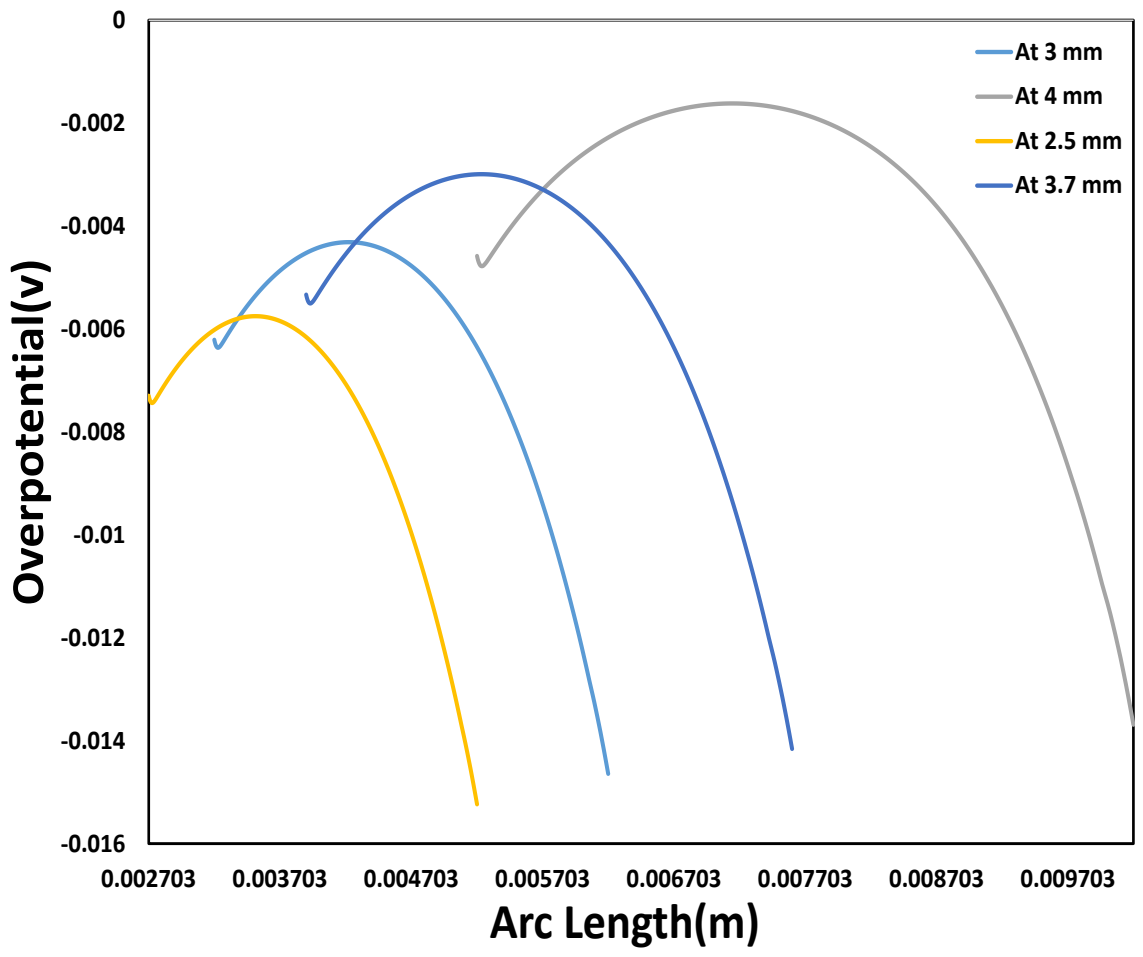

**Fig. 8** Overpotential distribution with compression of electrode.

The electrode was compressed while the electrolyte flow rate $v$ was held constant. After compression, the electrode's thickness is reduced from 4 mm to 2.5 mm. The electrode overpotential is calculated using the mechanism model's numerical analysis, as shown in Figure 8. In figure 8 it is seen that overpotential is higher at the beginning but gradually it decreases. With compression of electrode the overpotential decreases monotonously. Comparing the thickness reduced from 4 to 2.5 mm, the overpotential also decreased by 4mV. From this, it can be demonstrated that the enhancement in cell performance is mostly applicable to the decrease in ohmic loss induced by electrode compression. As this model is based on the state of charge, the drop of the electrode overpotential means that the performance of the battery is improved, the charging efficiency is improved, and the internal loss is reduced.

Figure 9 shows that under compression of the electrode the current density gradually decreases. At 4 mm the current density is the maximum and at 2.5 mm current density is minimum. From this, it can be seen that when the electrode is compressed, the current density decreases, resulting in the reduction of the electrode porosity and the increase of the specific surface area and conductivity. As a result, conductivity increases with the compression of the electrode.

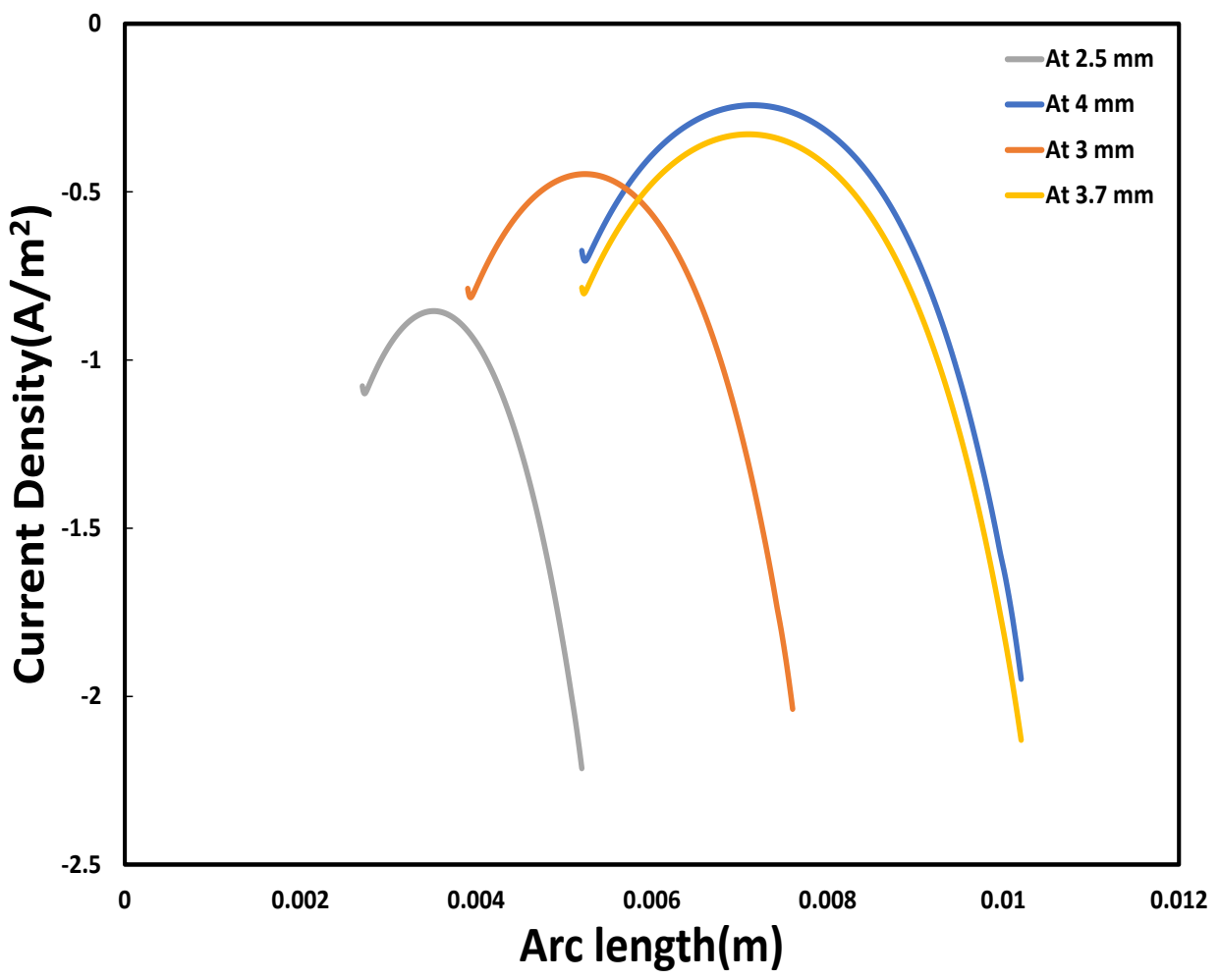

**Fig. 9** Current density distribution with compression of electrode.

In figure 10, the distribution of electrode potential decreased gradually with the compression of electrode. It can be found from the figure that terminal electrode potential decreases and because of the state of the charges the decrease of electrode potential means the battery performance is improved. For the 4 mm electrode, the potential is higher and for 3 mm, it decreases to a lower point. Electrode potential denotes the electron reduction and oxidation rate. As a result, electron transfer rate and current flow increase with compression of the electrode.

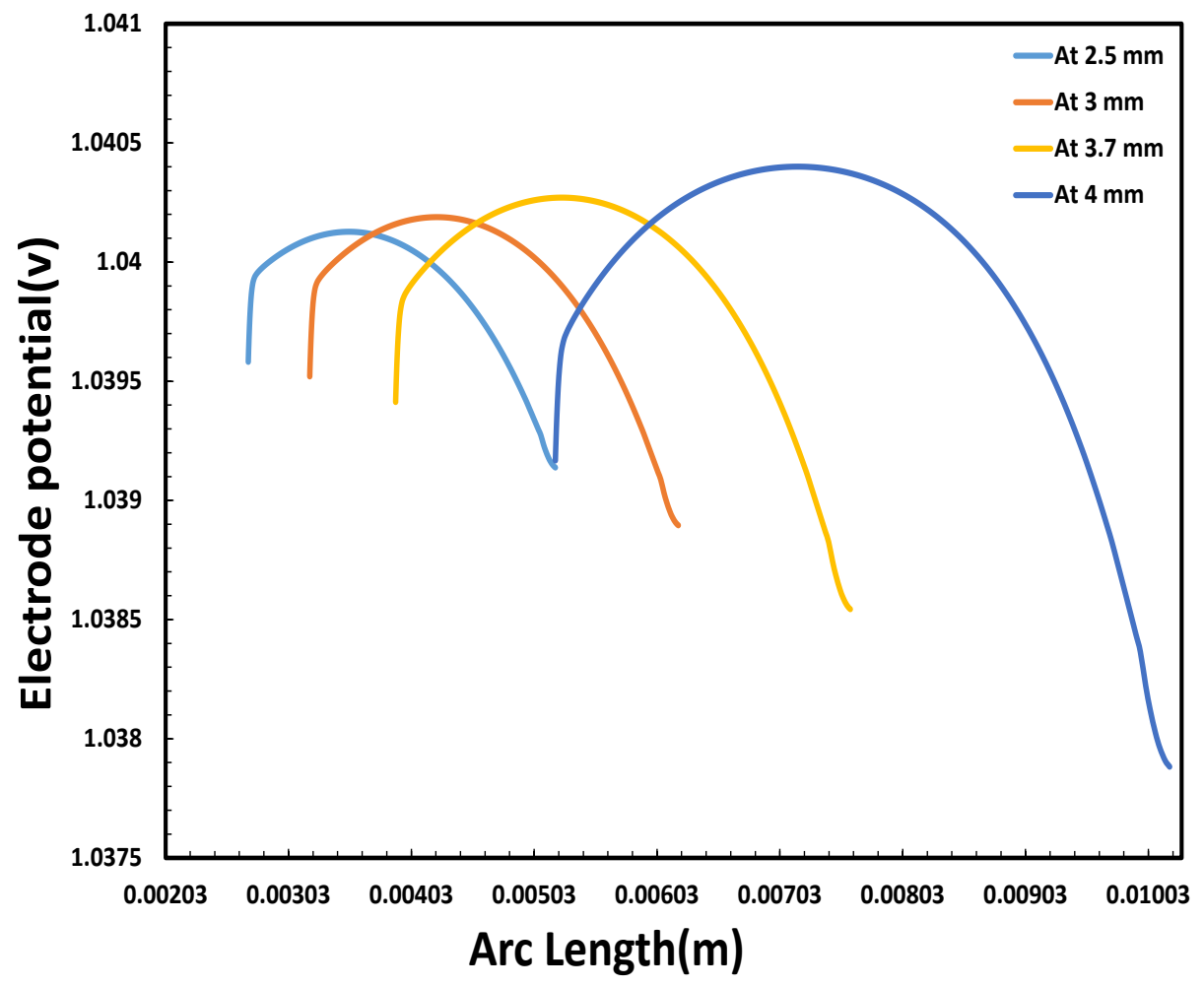

**Fig. 10** Electrode potential distribution with compression of electrode.

So, the performance of the battery is increased by the compression of the electrode.

## 5. Conclusion

NVRFB is modeled and simulated in this study. This simulation model was created to investigate the effects of dissociation rate, electrolyte flow rate, and electrode compression on performance attributes. Based on the dissociation rate of the Nickel Vanadium ions, it can be stated that more potential is required to activate the reactions with higher velocity and less potential is required with lower velocity. More reaction occurs at the membrane when the electrolyte flow rate is optimized. The porosity of the electrode decreases as the surface area, conductivity, and electron transfer rate increase. As ohmic loss decreases and current density increases, the overpotential decreases as well. As porosity decreases, so does the electrode potential over the surface. The oxidation-reduction rate increases as the electrode potential decreases. This signifies the battery's performance has been enhanced.